%% file: 3yearSolarWIMPPaper.tex
\documentclass[twocolumn,epjc3]{svjour3}

\usepackage[english]{babel}
\RequirePackage[T1]{fontenc}
\RequirePackage{graphicx}
\RequirePackage{mathptmx}      
\RequirePackage{flushend}
\RequirePackage[numbers,sort&compress]{natbib}
\RequirePackage[colorlinks,citecolor=blue,urlcolor=blue,linkcolor=blue]{hyperref}

\usepackage{booktabs}
\usepackage{pstricks}
\usepackage{calc}
\usepackage{amsmath}

\newcommand{\specialcell}[2][c]{\begin{tabular}[#1]{@{}c@{}}#2\end{tabular}}

\newcommand{\chb}{$b\bar{b}$}
\newcommand{\chW}{$W^{+}W^{-}$}
\newcommand{\chtau}{$\tau^{+}\tau^{-}$}

\newcommand{\mchtau}{\tau^{+}\tau^{-}}
\newcommand{\mExp}[1]{\times10^{#1}}
\newcommand{\unitdegree}{$^{\circ}$}
\newcommand{\UNIT}[1]{\,#1}

\journalname{Eur.~Phys.~J.~C}

\begin{document}
\title{Search for annihilating dark matter in the Sun with 3 years of IceCube data}
\onecolumn
\input{ICauthors.tex}
\date{Received: 20.12.2016 / Accepted: 08.02.2017 }
\maketitle

\titlerunning{Search for annihilating dark matter in the Sun}
\authorrunning{IceCube Collaboration}

\begin{abstract}
We present results from an analysis looking for dark matter annihilation in the Sun with the IceCube neutrino telescope. Gravitationally trapped dark matter in the Sun's core can annihilate into Standard Model particles making the Sun a source of \UNIT{GeV} neutrinos. IceCube is able to detect neutrinos with energies {$>$100\UNIT{GeV}} while its low-energy infill array DeepCore extends this to {$>$10\UNIT{GeV}}. This analysis uses data gathered in the austral winters between May 2011 and May 2014, corresponding to 532 days of livetime when the Sun, being below the horizon, is a source of up-going neutrino events, easiest to discriminate against the dominant background of atmospheric muons. The sensitivity is a factor of two to four better than previous searches due to additional statistics and improved analysis methods involving better background rejection and reconstructions. The resultant upper limits on the spin-dependent dark matter-proton scattering cross section reach down to {$1.46\mExp{-5}$\UNIT{pb}} for a dark matter particle of mass {500\UNIT{GeV}} annihilating exclusively into \chtau particles. These are currently the most stringent limits on the spin-dependent dark matter-proton scattering cross section for WIMP masses above {50\UNIT{GeV}}. 

\end{abstract}

\twocolumn
\section{Introduction}
Astrophysical observations provide strong evidence for the existence of dark matter (DM). However its nature and possible particle constituents remain unknown. Interesting and experimentally accessible candidates are the so called `Weakly Interacting Massive Particles (WIMPs)'  - expected to exist in the mass range of a few GeVs to a few TeVs (see~\cite{HooperPDM} for a comprehensive review). If DM consists of WIMPs, they can be gravitationally captured by the Sun~\cite{Press:1985ug, Gaisser:1986ha, Srednicki:1986vj, Ritz:1987mh}, eventually sinking to its core, where they may pair-annihilate into standard model particles producing neutrinos. Given enough time, the capture and annihilation processes would reach an equilibrium~\cite{WIMPAnni} with, on average, only as many DM particles annihilating as are captured per unit time. This DM-generated neutrino flux may be detected at terrestrial neutrino detectors such as IceCube. As the region at the center of the Sun where most of the annihilations will occur is very small, the search is equivalent to looking for a point-like source of neutrinos. Neutrinos above {1\UNIT{TeV}} have interaction lengths significantly smaller than the radius of the Sun and are mostly absorbed. As a result all the signal is expected in the range of a few GeVs to {$\sim$1\UNIT{TeV}}.

IceCube (section~\ref{sec:detector}) detects neutrinos by looking for the Cherenkov light from charged particles produced in the neutrino interactions. While charged-current (CC) interactions of $\nu_{\mu}$ (and $\bar{\nu}_{\mu}$) produce muons that traverse the detector producing clear track-like signatures, the vast majority of such events observed by IceCube are muons produced when cosmic rays interact in the upper atmosphere (section~\ref{sec:sim}). Although they are observed only in the downgoing direction as they do not cross the Earth, their dominance in numbers by five orders of magnitude with respect to the atmospheric neutrino flux require strong measures for their rejection. Similar events created by the interactions of atmospheric neutrinos in ice are, except for their spectral composition, indistinguishable from neutrino events of extra-terrestrial origin and so remain an irreducible background. A correctly reconstructed up-going event thus must come from a neutrino interaction. This analysis focuses exclusively on these track-like upgoing events. At the energies relevant to this analysis, the direction of the muon serves as a proxy for the direction of the initial neutrino and allows us to identify a directional excess from the Sun in reconstructed events. 

We exploit this fact in the event selection (section~\ref{sec:eventsel}) for this analysis, using only seasons where the Sun is a source of up-going signal events. Furthermore we devise an event selection which minimizes atmospheric muon background contamination and limits the impact of mis-reconstructed events. The remaining samples of events are then analyzed using an unbinned maximum likelihood ratio method~\cite{Method}, looking for an excess of events from the direction of the Sun. This method compares the observed angles and energy spectrum to signal expectations from different simulated WIMP masses and annihilation channels (section~\ref{sec:anamethod}). Section~\ref{sec:results} and \ref{sec:concl} present the results of this analysis as well as their interpretation in the framework of the larger effort to detect dark matter.

\begin{figure*}[]\centering
  \includegraphics[width=0.5\linewidth]{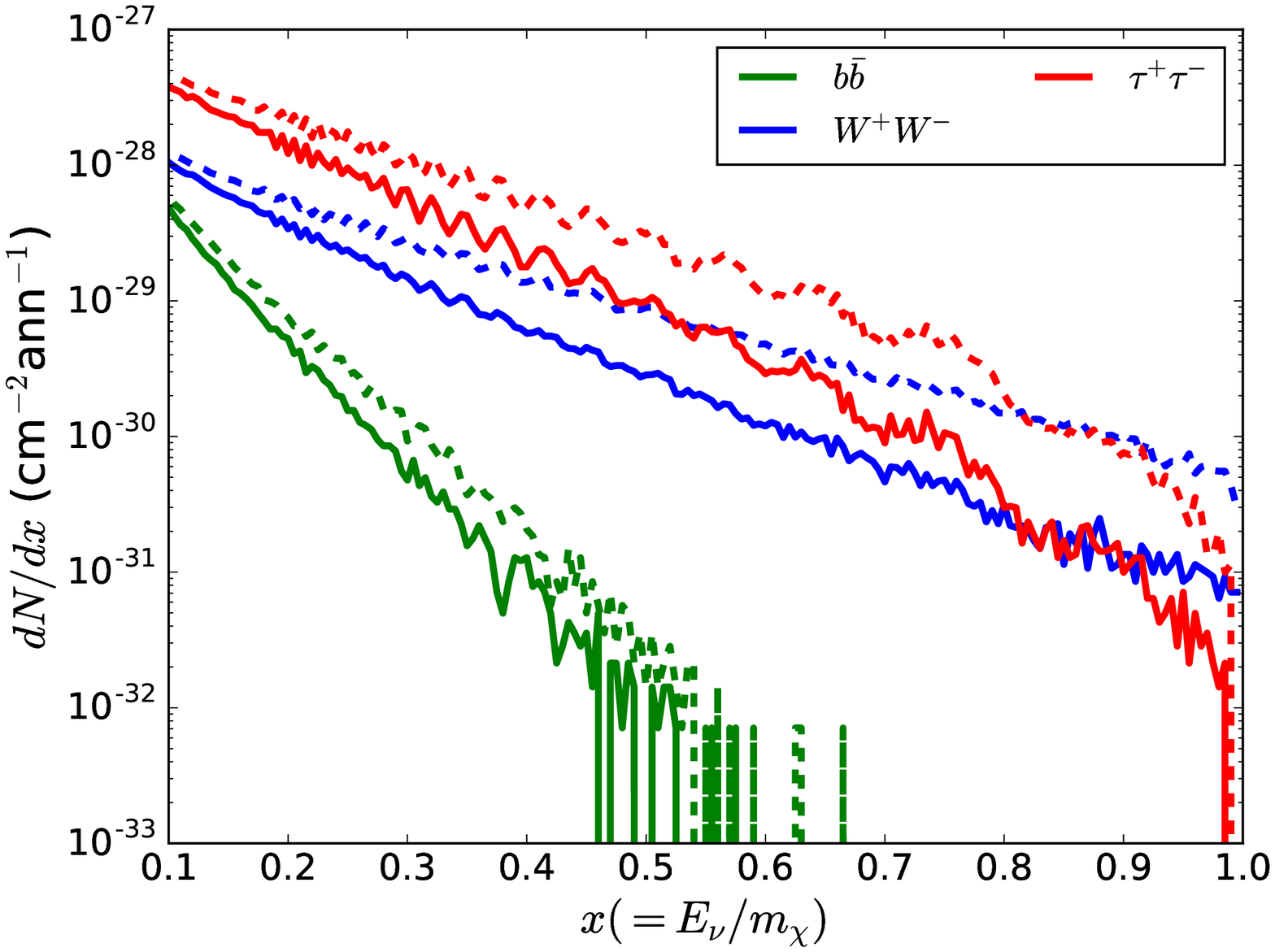}\includegraphics[width=0.5\linewidth]{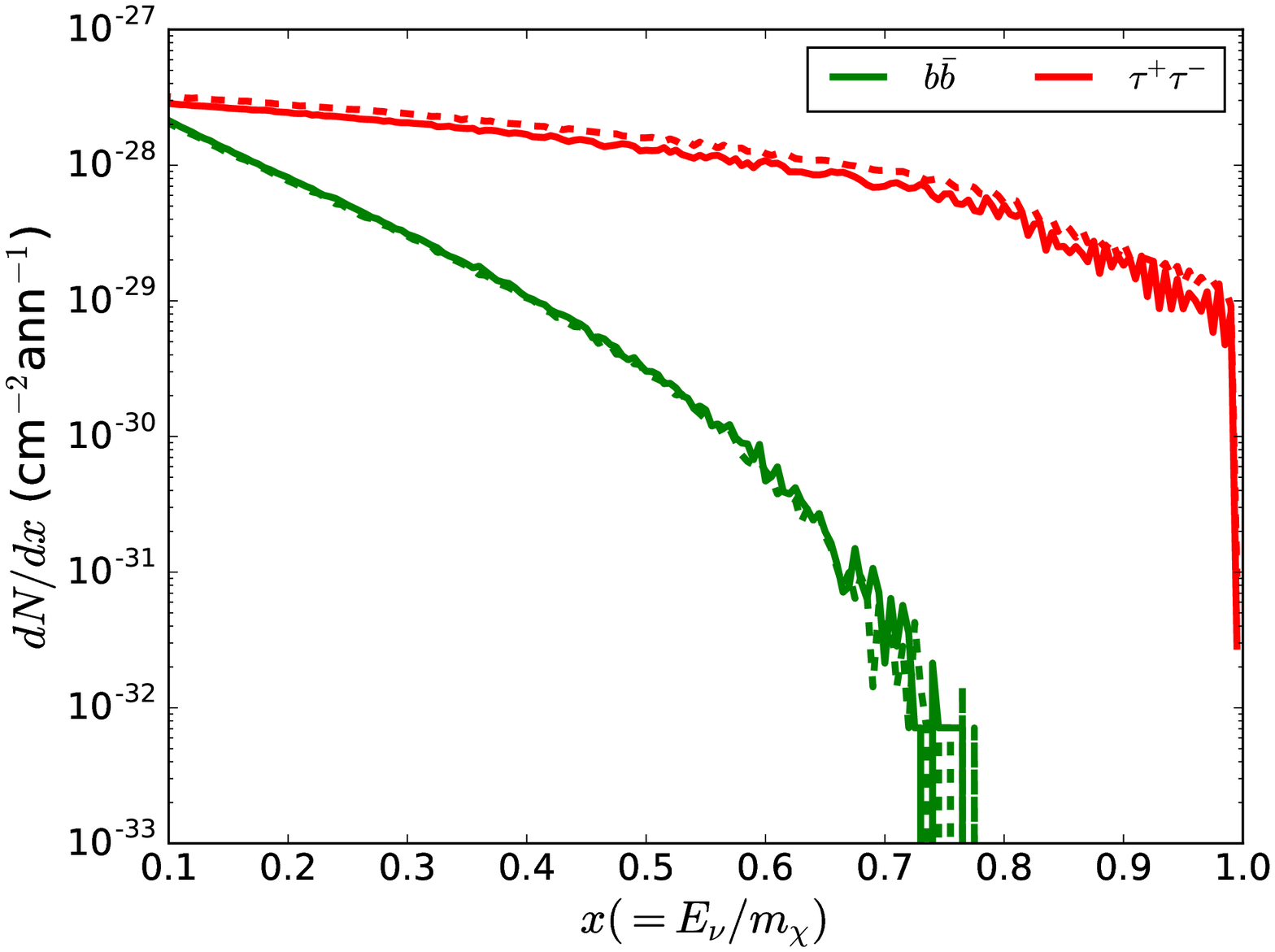}
  \caption{Differential $\nu_{\mu}$(solid) and $\bar{\nu}_{\mu}$(dashed) fluxes at Earth from the annihilations of {1\UNIT{TeV}} (left) and {50\UNIT{GeV}} (right) WIMPs in the Sun respectively, including absorption and neutrino oscillation effects (visible as wiggles in the plot on the left), as predicted by \texttt{WimpSim}~\cite{WIMPSim}. The $\bar{\nu}_{\mu}$ fluxes are higher than the $\nu_{\mu}$ fluxes at lower energies since their interactions with the matter of the Sun are helicity suppressed.}
  \label{fig:Fluxes}
\end{figure*}

\section{The Detector}\label{sec:detector}
IceCube is a cubic-kilometer neutrino detector installed in the ice~\cite{IcePaper} at the geographic South Pole~\cite{FirstYearPerformancePaper} between depths of 1450\,m and 2450\,m. Neutrino reconstruction relies on the optical detection of Cherenkov radiation emitted by secondary particles produced in neutrino interactions in the ice or the nearby bedrock. The photons are detected by photomultiplier tubes (PMT)~\cite{PMTPaper} housed in Digital Optical Modules (DOM)~\cite{DOMPaper}. Construction of the detector started in 2005 and the detector has been running in its complete configuration since May 2011, with a total of 86 strings deployed, each equipped with 60 DOMs.

The principal IceCube array consists of 78 strings ordered in a hexagonal grid with a string spacing of approximately {125\UNIT{m}}, an inter-DOM spacing of {17\UNIT{m}} along each string, and can detect events with energies as low as {$\sim$100\UNIT{GeV}}. Eight infill strings are deployed in the central region of IceCube to form DeepCore, optimized in geometry and instrumentation for the detection of neutrinos at further lower energies, down to {$\sim$10\UNIT{GeV}}. A layer of dust, causing a region of increased scattering and absorption, intersects the detector at depths between 1860\UNIT{m} and 2100\UNIT{m}. Since the ice becomes more transparent at increasing depth, the main part of the DeepCore instrumentation is deployed below the dust layer with an inter-DOM spacing of only 7\,m. A veto cap of additional 10 DOMs deployed above the dust-layer completes the DeepCore strings. A majority of the DeepCore DOMs are equipped with PMTs of higher quantum efficiency to increase light collection. These DeepCore strings, along with the seven adjacent standard IceCube strings, constitute the fiducial region of the DeepCore subarray for the purpose of this analysis~\cite{DeepCorePaper}. For DM annihilations producing neutrinos above {$\sim$100\UNIT{GeV}}, the full instrumented volume of the principal IceCube array contributes to the sensitivity, while for lower DM masses when the signal neutrinos are below the IceCube threshold, only the DeepCore fiducial volume is relevant. The IceCube array nevertheless plays a role in identifying and rejecting background events at these lower energies.

\section{Signal and background simulations}\label{sec:sim}
Neutrino flux predictions at Earth from WIMP annihilations in the Sun have been widely studied, for example in Ref.~\cite{WIMPSim}. We use the flux predictions from \texttt{DarkSUSY}~\cite{DarkSuSy} and \texttt{WimpSim}~\cite{WIMPSim} to simulate signals for the IceCube detector according to specific annihilation scenarios, incorporating effects from absorption in the Sun as well as neutrino oscillations\cite{OscillationProperties}. Events from all three flavours of signal neutrinos are simulated. When WIMPs annihilate into \chW (see Fig.~\ref{fig:Fluxes}), the $W$ bosons decay promptly and neutrino emission from the leptonic decay channels peaks at energies close to the mass of the WIMP. The \chtau channel produces a similar distribution of neutrinos in energy with a higher overall normalization. These are referred to as `hard' channels. When the WIMP annihilates predominantly to a `soft' channel such as \chb, the neutrino emission peaks at energies much below the mass of the WIMP, since the $b$ quarks hadronize before they can decay to produce neutrinos. \texttt{WimpSim} does not account for modifications to the spectrum originating from the radiation of electroweak gauge bosons by the intermediate and final states of the decay process. These effects have been studied in \cite{PPPC4DMnu}. Since both {W$^{\pm}$} and Z bosons decay promptly to produce high energy neutrinos, the net effect of these electroweak corrections is to harden the fluxes from the softer channels and enhance signal rate expectations.

The principal background of muons generated in the interactions of cosmic rays with the Earth's atmosphere is simulated using the \texttt{CORSIKA} package~\cite{CORSIKA}. Atmospheric neutrino interactions with the ice and the bedrock surrounding the detector are simulated using neutrino-generator (\texttt{NuGen})~\cite{Nugen} above {150\UNIT{GeV}} with the cross sections of~\cite{SubirCS} and \texttt{GENIE}~\cite{genie} below {150\UNIT{GeV}}. The atmospheric neutrino flux predictions of~\cite{Honda2007} are used to weight \texttt{NuGen} and \texttt{GENIE} simulated datasets to validate the data processing and event selection.

\section{Event Selection}\label{sec:eventsel}
\begin{figure}[h] 
  \centering 
  \begin{minipage}[t][][t]{\linewidth}
    \begin{minipage}[t][][t]{0.49\linewidth}
        \centering 
        \includegraphics[scale=0.3]{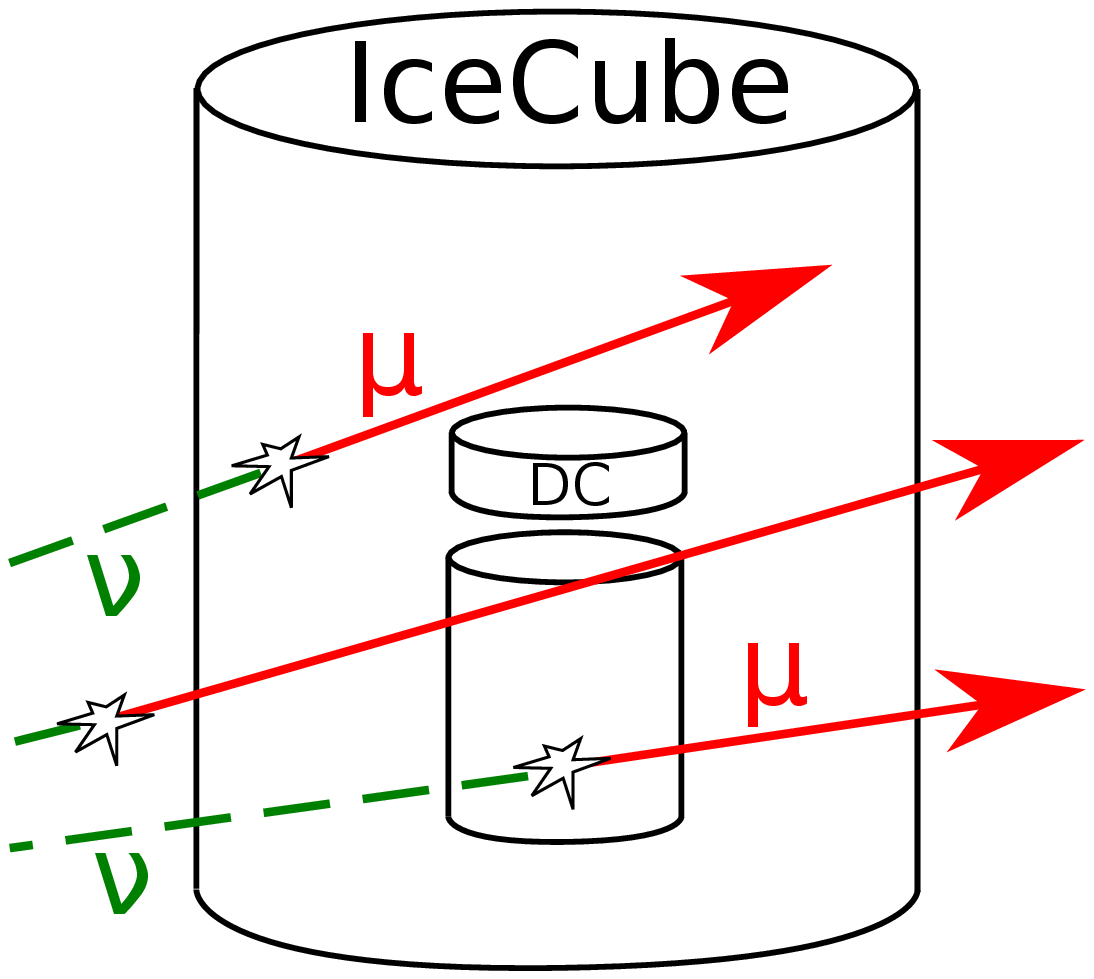}\\
        a) IceCube selection
    \end{minipage}
    \begin{minipage}[t][][t]{0.49\linewidth}
        \centering 
        \includegraphics[scale=0.3]{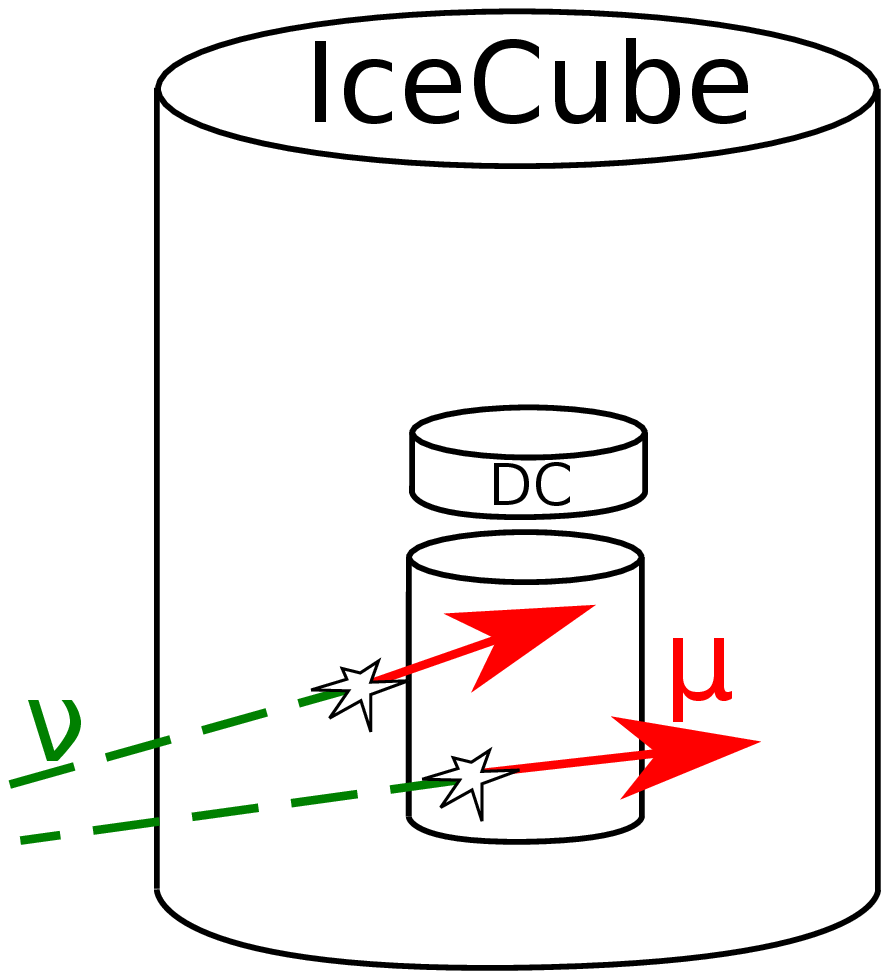}\\
        b) DeepCore selection
    \end{minipage}
  \end{minipage}
  \caption{The two event selection strategies for this analysis. The IceCube dominated high energy sample (a) is sensitive to neutrinos above {$\sim$100\UNIT{GeV}}. Most of the sensitivity for neutrino signals below {100\UNIT{GeV}} comes from the DeepCore (DC) dominated low energy sample (b). This approach is similar to that of earlier IceCube analyses~\cite{MatthiasPaper}.}
  \label{fig:Evselections}
\end{figure}

The energy range of the expected signal (a few TeV at maximum) and the event topologies in the detector at these energies dictate the event selection strategies. For WIMP masses less than {200\UNIT{GeV}}, which produce signal neutrinos mostly with energies below the IceCube threshold, only DeepCore will contribute significantly towards the effective volume. However, for higher WIMP masses, where a significant fraction of the resultant neutrinos are above the IceCube threshold, the full instrumented volume of IceCube comes into play. Consequently we select two non overlapping samples of events as illustrated in Fig.~\ref{fig:Evselections}.

To optimize the event selections for the analysis, we consider two scenarios: WIMPs annihilating completely into \chW and WIMPs annihilating completely into \chb. For WIMP masses below {80.4\UNIT{GeV}}, the mass of the $W$ boson, we consider the WIMP annihilating into \chtau, since annihilations to \chW are not kinematically allowed. Since the detector acceptance is energy dependent, cuts have to be optimized for the spectral composition of the expected signal flux. 

\begin{figure}[]\centering
  \includegraphics[width=1.0\linewidth]{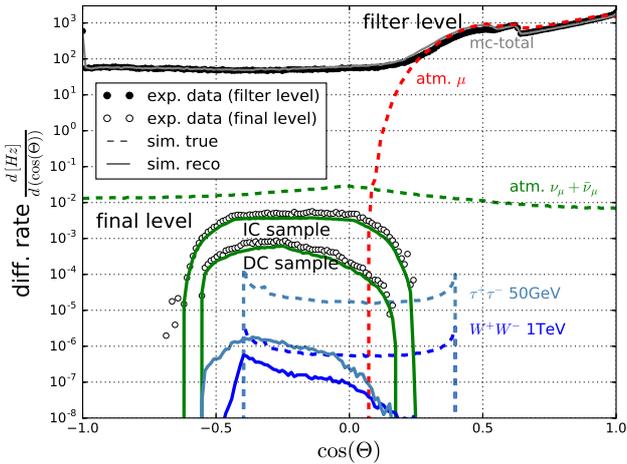}
  \caption{Zenith distributions for simulation, indicated for their simulated particle direction (MC, dashed lines) and reconstructed direction (reco, solid lines), and data, with only reconstructed directions (circles), at filter level (L2) and analysis level. At filter level the down-going atmospheric $\mu$-background (red dashed), dominates even the up-going region for the recorded data (solid circles), because of false direction reconstructions (solid grey). The flux expectation of atmospheric $\nu_{\mu}$ (green dashed) are indicated~\cite{Honda2007}. After removal of background events in the event selection reconstructed track-like atmospheric $\nu_{\mu}$-events (green solid) dominate the remaining exp.\ data (open circles) at final level. The plot also shows the obtained limits on the solar WIMP $\nu_{\mu}$ signal flux obtained by this analysis for two different WIMP models, which are reconstructed in DeepCore ({50\UNIT{GeV}} \chtau, light blue) and IceCube ({1\UNIT{TeV}} \chW, dark blue) at analysis level (solid) and scaled by their selection efficiency at filter level (dashed).}
  \label{fig:SolarEvSelectionOutlook}
\end{figure}

Within IceCube, a standard set of filters pre-select signal-like events and reduce the rate of the dominant background of atmospheric muons, subsequent to which reconstructions specific to the event topology are carried out, at what is known as the filter level or level 2 (L2). We focus on a stream of data from three of these filters, a low-energy event filter on the topological region of DeepCore and two further filters selecting muon-like events in the bigger IceCube array. One of these filters favours short low energy upward going tracks. The other selects general bright track-like events, both up and down-going, where the latter class is restricted to events starting within the detector. After these filters the data rate is reduced from {3\UNIT{kHz}} to about {100\UNIT{Hz}}. Still, atmospheric muons constitute the overwhelming majority of events. At this stage, about 30\% of the neutrino events recorded by IceCube include a coincident atmospheric muon event. The goal is to further reduce the data with a series of reconstructions and cuts to a sample of signal-like neutrino events This sample will be, however comprised almost exclusively of atmospheric neutrino events, an irreducible background to the analysis. Fig.~\ref{fig:SolarEvSelectionOutlook} provides a comparative summary of the event rates at filter and analysis level.

\subsection{Data Treatment} 
The processing of IceCube data proceeds in sequential steps, referred to as selection levels. It involves the abstraction of the recorded analog to digital converter data as photons impacting on single PMTs (hits), the removal of nuisance hits caused by detector noise and coincident events\footnote{two or more events being present in the detector at the same time and ending up in the same readout window. An effect observed in {$\sim$10\%} of recorded events, up to {30\%} depending on filter stream selection.}, event reconstructions of increasing complexity and event selection cuts. The reconstructions assume single event topologies built up only by hits that are caused by the radiating particle. They can easily be misled by nuisance hits, making hit cleaning a priority for any IceCube analysis.

This analysis makes use of a new approach for the necessary noise cleaning and separation of coincident events by an agglomerative hit clustering algorithm~\cite{MarcelProceeding,MarcelThesis}. It operates progressively on the IceCube data stream described by the time-distribution of hits. Within the algorithm, which takes into account the hexagonal design of the detector and the difference in instrumentation density between its components, the physical causal relation between consecutive hits is analyzed. If found to be causally connected, hits are considered to form a cluster. Clusters grow by further addition of more connected hits, while unconnected hits are rejected. Each such identified cluster can later be attributed to a particle (sub)event within the detector. Persistent errors, such as the splitting of a single event into two separate subevents are corrected by a subsequent algorithm described in~\cite{MarcelProceeding}, which probes the recombination of subevents back into a single event. The combination of these algorithms performs {50\%} better than previous approaches, in both selecting the correct hits created by the radiating particle as well as the correct separation of events arriving in coincidence.

\subsection{IceCube Event Selection}
From the {$\sim$100\UNIT{Hz}} of data from the three filters at L2, cuts favoring horizontal, well reconstructed events are used to select {$\sim$3\UNIT{Hz}} of data (L3). The position of the Sun varies between $\sim66^{\circ}$ and $104^{\circ}$ in zenith angle. Consequently the signal events are expected to be horizontal within the detector. Subsequently, events that have more hits outside the DeepCore fiducial volume or at least 7 hits in the IceCube strings are selected. More sophisticated and computationally intensive reconstructions are performed at this stage. A Bayesian likelihood-based reconstruction that uses the prior knowledge that the data are still dominated by down-going muons is used, along with consistency tests between the various track reconstructions performed so far. This reduces the data rate to {$\sim$140\UNIT{mHz}} (L4). Subsequently, a Boosted~\cite{AdaBoost} Decision Tree (BDT) is used to quantify each event as signal or background-like using a score, based on a set of variables describing the event topology and direction, as well as relative positions and arrival times of the various photon hits within the detector. The BDT is trained on simulated signal events of the \chW-annihilation channel of {1\UNIT{TeV}} WIMPs.

The optimum threshold on the BDT score was determined using the Model Rejection Factor method described in~\cite{GaryHillMRFmethod} for the same signal hypothesis. The remaining {$\sim$2.9\UNIT{mHz}} of data (L5) are dominated by up-going muons from charged current interactions of atmospheric $\nu_{\mu}$ (and $\bar{\nu}_{\mu}$). The angular resolution of this sample is further improved using a reconstruction which utilizes tabulated photon arrival time distributions obtained from simulation as described in~\cite{PhotoSplinePaper}. The median neutrino angular resolution for this final sample ranges from {$\sim$6\unitdegree} for a {100\UNIT{GeV}} neutrino to {$<$1\unitdegree} for a {1\UNIT{TeV}} neutrino.

\begin{table}[]
  \centering
  \caption[ICHighEn rate summary]{Rate summary for the IceCube event selection. The signal efficiencies are with respect to L2 for the {1\UNIT{TeV}}$\to W^{+}W^{-}$ signal. The atmospheric muon neutrino rates indicate the sum of $\nu_{\mu}$ and $\bar{\nu}_{\mu}$ in the expected ratio. The discrepancy between the data rate at L2 and the total Monte Carlo rate is due to deficiencies in \texttt{CORSIKA}. As cuts reject most of the atmospheric muon background, the discrepancy becomes smaller. The final analysis method uses randomized data to estimate the background, and is not affected by this discrepancy.}
  \label{tab:ratesummaryICHighEn}
    \begin{tabular}{|c|c|c|c|c|c|c|}
    \toprule
    \specialcell{Cut\\Level} & \specialcell{Data\\(Hz)}  & \specialcell{\texttt{CORSIKA}\\(Hz)} &\specialcell{Atmos $\nu_{\mu}$\\(Hz)}  & \specialcell{Sig eff \\(\%)}\\
    \midrule
    L2 & 98.4  & 72.8 & 18.6$\mExp{-3}$ & 100 \\
    L3 & 2.81 & 3.32 & 8.4$\mExp{-3}$& 82 \\
    L4 & 0.14 & 0.15  & 5.2$\mExp{-3}$& 63 \\
    L5 & 2.9$\mExp{-3}$ &  0.8$\mExp{-3}$& 2.1$\mExp{-3}$& 37\\
    \bottomrule
  \end{tabular}
\end{table}

\subsection{DeepCore Event Selection}
The fiducial region of DeepCore is already embedded deep within the detector. In consequence, using a selection of DeepCore dominated events with less than 7 hits on regular IceCube strings (the compliment of the selection criterion for the IceCube event selection) already provides a certain degree of background rejection via containment and starting requirement for events. These two properties are further exploited for the identification of events originating from neutrino interactions.

The DeepCore event selection starts with events selected by any filter at L2. Straight cuts which enforce minimal event quality and a loose selection for low energy horizontal events are applied. To reject down-going events still contained in the sample, hit-based vetos requiring no hits in the outer and top-most DOMs are applied. Subsequently, events are reconstructed with the reconstruction described in~\cite{PhotoSplinePaper} followed by further straight cuts, which reduce the content of atmospheric muons within the sample (L3+L4). After this a BDT, which is trained on the selection of track-like $\nu_{\mu}$-events by variables expressing the position, incoming direction and reconstruction quality of events, is used. Thereafter a cut is applied requiring the zenith angle to be reconstructed within {10\unitdegree} of the actual Sun position. A second BDT, which is trained on the exact signal properties by additional energy-sensitive event variables, is used to further refine the event classification (L6). A loose cut on the second BDT-score sufficiently reduces the sample, so that the computationally demanding energy reconstruction described in~\cite{3yearoscillation} can be applied to all remaining events (L7). This reconstruction estimates the total energy of the incoming neutrino from the length of the muon track as well as the photons from hadronic debris from the charged current interaction, when the interaction has taken place within the instrumented detector volume. The sample now contains all variables needed for the likelihood analysis procedure. The BDT-score cut can be further optimized for the best sensitivity for a broad range of WIMP models at the low WIMP mass end and obtain the sample at L8.

\begin{figure*}[]\centering
  \includegraphics[width=0.50\linewidth]{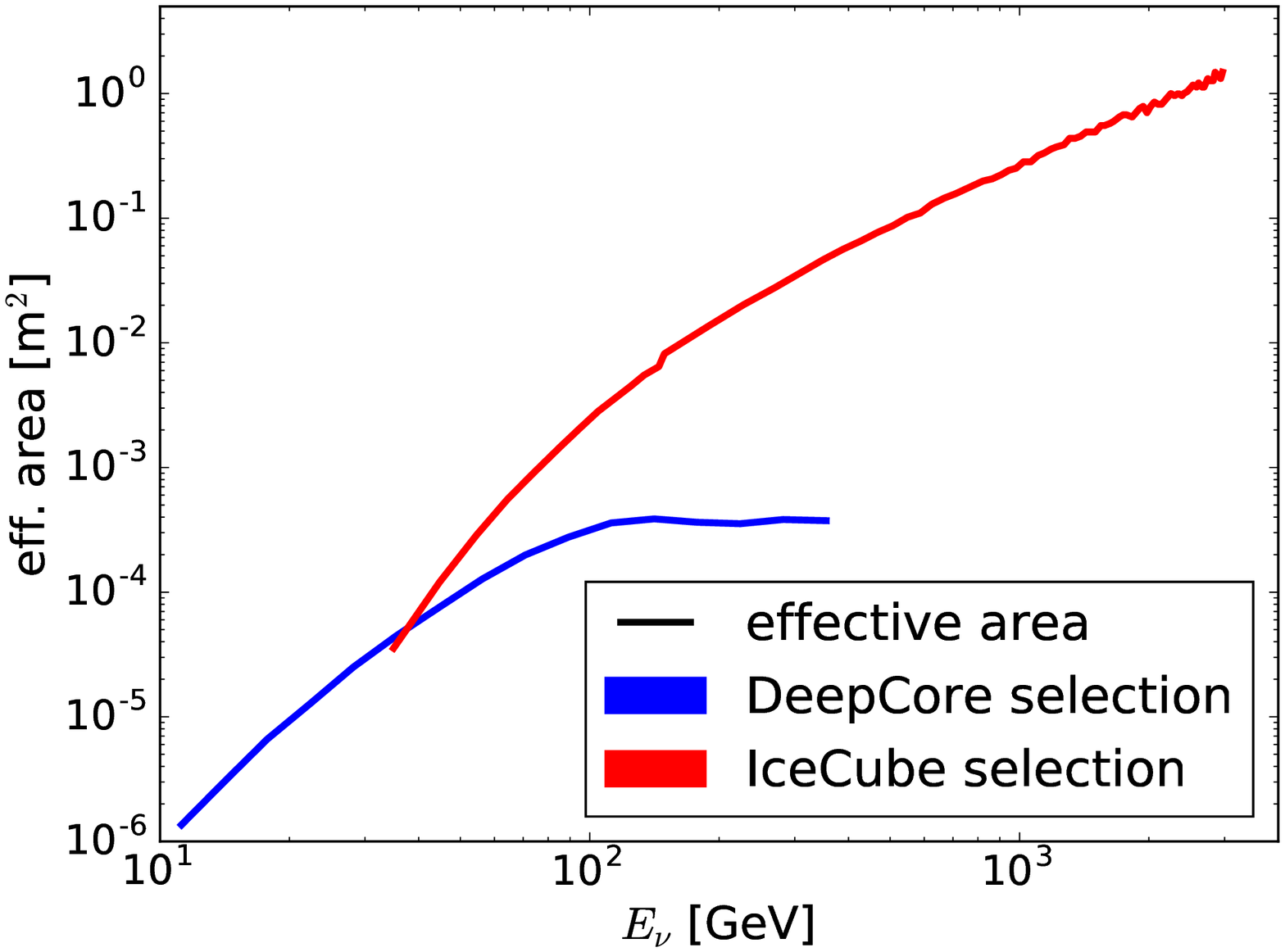}\includegraphics[width=0.50\linewidth]{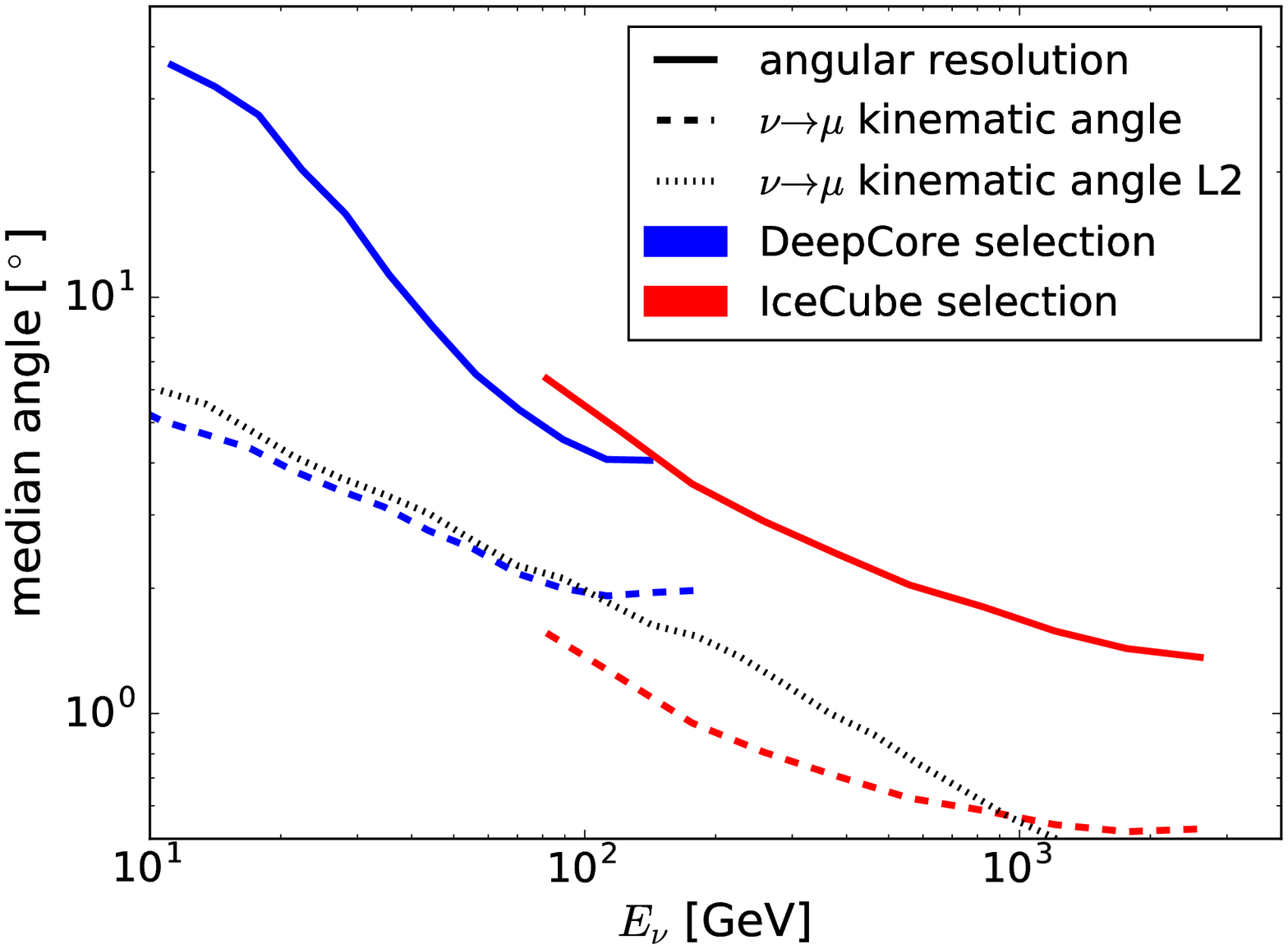}
  \caption{Event selection performance. Left: $\nu_{\mu}$ + $\bar{\nu}_{\mu}$ effective area, derived from Monte Carlo simulations performed using \texttt{GENIE}~\cite{genie} for the DeepCore selection and \texttt{NuGen}~\cite{Nugen} employing the cross sections as calculated by~\cite{SubirCS} for the IceCube selection.  Right: median angular resolution as a function of true neutrino energy. The dashed lines indicate the median kinematic angle between the incoming neutrino and the muon. Events from CC interactions of neutrinos with energies higher than {$\sim$100\UNIT{GeV}} are preferentially included in the IceCube selection as the range of the muon is higher than the containment requirements of the DeepCore selection. Those that are included in the DeepCore selection are the ones of lower energy and those in which a larger fraction of the neutrino energy has been transferred to the hadronic cascade. For the latter events, a larger fraction of the observed photon yield comes from the hadronic cascade, affecting the performance of the track direction reconstruction. In addition the kinematic scattering angle also is higher for such events. Consequently, a saturation effect can be seen in both angular resolution and kinematic angle lines for the DeepCore Selection, at energies of {$\sim$100\UNIT{GeV}}.}.
  \label{fig:EvselectionsPerf}
\end{figure*}

\paragraph*{}
The selection criteria described in the two sections above are applied to data from the austral winters between May 2011 and March 2014. This produces two non overlapping samples with the $\nu_{\mu} + \bar{\nu}_{\mu}$ effective areas and angular resolutions shown in Fig.~\ref{fig:EvselectionsPerf}, corresponding to 532 days of operation of IceCube-DeepCore. While this Tables~\ref{tab:ratesummaryICHighEn} and \ref{tab:ratesummaryDCLow} summarize the rates and neutrino purities of the two streams at various levels of the event selection.

During the austral summer, when the Sun is above the horizon and a source of down-going neutrinos, an additional background of down-going atmospheric muons, {$\sim10^{5}$} higher in rate than atmospheric neutrinos at filter level, dominates over the signal. For this data taking period, in order to reach a sample of suitable events for analysis, considerably harder cuts are required, diminishing the acceptance of neutrino events. Samples isolated from these periods of operation of IceCube-DeepCore~\cite{MarcelProceeding, GenevaICRC2015} have been found to not contribute significantly to the sensitivity and are thus not further considered.

\begin{table}[]
  \centering
  \caption[DCHighEn rate summary]{Rate summary for the DeepCore event selection. The signal efficiencies are with respect to L2 for the {50\UNIT{GeV}}$\to \mchtau$ -- signal.}
  \label{tab:ratesummaryDCLow}
  \begin{tabular}{|c|c|c|c|c|c|}
    \toprule
    \specialcell{Cut\\Level} & \specialcell{Data\\(Hz)} & \specialcell{\texttt{CORSIKA}\\(Hz)} & \specialcell{Atmos $\nu_{\mu}$\\(Hz)} & \specialcell{Sig eff \\(\%)}\\
    \midrule
    L2 & 25.6 & 25.1 & 5.90 $\mExp{-3}$ & 100 \\
    L3 & 1.37 & 1.03 & 3.36$\mExp{-3}$ & 58 \\
    L4 & 0.74 & 0.66 & 3.24$\mExp{-3}$ & 57 \\
    L5 & 0.55 & 0.48 & 2.48$\mExp{-3}$ & 42 \\
    L6 & 66.9$\mExp{-3}$ & 59.6$\mExp{-3}$ & 2.140$\mExp{-3}$ & 38\\
    L7 & 1.82$\mExp{-3}$ & 2.07$\mExp{-3}$ & 0.423$\mExp{-3}$ & 16\\
    L8 & 0.334$\mExp{-3}$ & 0.143$\mExp{-3}$ & 0.220$\mExp{-3}$ & 10\\
    \bottomrule
  \end{tabular}
\end{table}

\section{Analysis Method}\label{sec:anamethod}

An unbinned maximum likelihood ratio method~\cite{Method} is subsequently used to look for a statistically significant excess of events from the direction of the Sun. The signal probability density function (p.d.f.), explicitly dependent on the event's reconstructed direction, $\vec{x}_{i}$, energy, $E_{i}$, and observation time, $t_{i}$, is given by:
\begin{equation} \label{eq:p.d.f}
  S_{i}(\vec{x}_{i},t_{i},E_{i},m_{\chi},c_{\chi}) =  \mathcal{K}(|\vec{x}_{i}-\vec{x}_{\odot}(t_{i})|, \kappa_{i}) \times \mathcal{E}_{m_{\chi},c_{\chi}}(E_{i}),
\end{equation}
\noindent where $\mathcal{K}$ stands for the spatial and $\mathcal{E}$ for the spectral parts of the {p.d.f.} and $m_{\chi}$ and $c_{\chi}$ stand for the mass and annihilation channel of the WIMP respectively. Here $\mathcal{K}$ is approximated by the monovariate Fisher-Bingham distribution~\cite{Fishbingh} from directional statistics, dependent on the opening angle, $\theta$, between the event and the direction of the Sun at observation time, $\vec{x}_{\odot}(t_{i})$, given by
\begin{equation} \label{eq:spacesun}
  \mathcal{K}(|\vec{x}_{i}-\vec{x}_{\odot}(t_{i})|, \kappa_{i}) = \frac{\kappa_{i} e^{\kappa_{i} \cos(\theta_{|\vec{x}_{i}-\vec{x}_{\odot}(t_{i})|})}}{2\pi(e^{\kappa_{i}} - e^{-\kappa_{i}})}
\end{equation}
\noindent The concentration factor $\kappa_{i}$ of the event $i$ is obtained from the likelihood-based estimate of the angular resolution of the track reconstruction~\cite{paraboloidpaper}. The energy part of the signal p.d.f.\ is constructed from signal simulations.

The background p.d.f.\ is:
\begin{equation} \label{eq:bkgpdf2}
  B_{i}(\vec{x}_{i},E_{i}) = D(\delta_{i})\times P(E_{i}|\phi_{\mathrm{atm}})
\end{equation}
where $D(\delta_{i})$ is the declination dependence and $P(E|\phi_{\mathrm{atm}})$ indicates the distribution of the energy estimator $E$ in the event sample which is constructed from the data dominated by atmospheric neutrinos.

\begin{figure}[]
  \centering
  \includegraphics[width=1.0\linewidth]{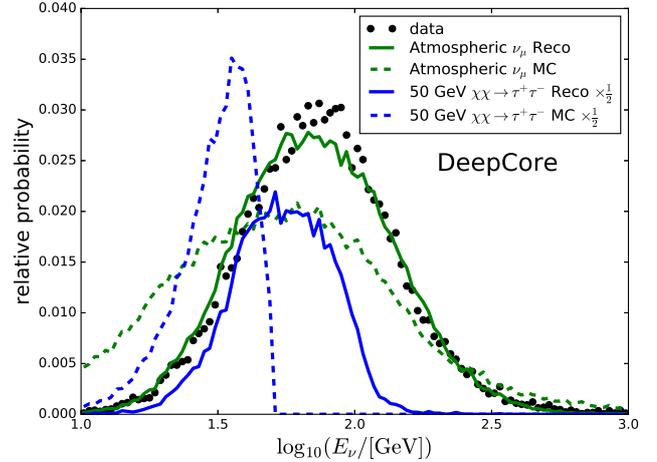}
  \caption{Distribution of the reconstructed energy (Reco) of the events for the DeepCore selection. For background and signal simulations, the true energy distribution (MC) is shown in dashed lines. The distributions for the signal are scaled down by a factor of $1/2$ for improved visualization.}
  \label{fig:EnergyPDF}
\end{figure}

The energy part of the signal and background p.d.f.s are the distributions of reconstructed energy obtained from signal simulations and observed data, respectively, and are used only for the DeepCore sample. They are illustrated in Fig.~\ref{fig:EnergyPDF}.

For a sample of $N$ events consisting of $n_\mathrm{s}$ signal events from the Sun and $N-n_{\mathrm{s}}$ background events, the likelihood can then be written as:
\begin{equation} \label{eq:likelihood2}
  \mathcal{L}(n_\mathrm{s}) = \prod\limits_{N}\left(\frac{n_\mathrm{s}}{N} S_{i} + (1-\frac{n_\mathrm{s}}{N}) B_{i}\right)
\end{equation}
The best estimate for the number of signal events in the sample is obtained by maximizing the likelihood ratio as defined in Ref.~\cite{Method}. The significance of the observation can be estimated without depending on Monte Carlo simulations by repeating the process on datasets randomized in right ascension. As the two event selections have no events in common, they can be combined statistically using the method described in Ref.~\cite{3yearPSPaper}. Confidence intervals on the number of signal events present within the sample are constructed using the method of~\cite{Feldman}.

\subsection{Systematic Uncertainties}
\begin{figure*}[]\centering
  \includegraphics[width=0.8\linewidth]{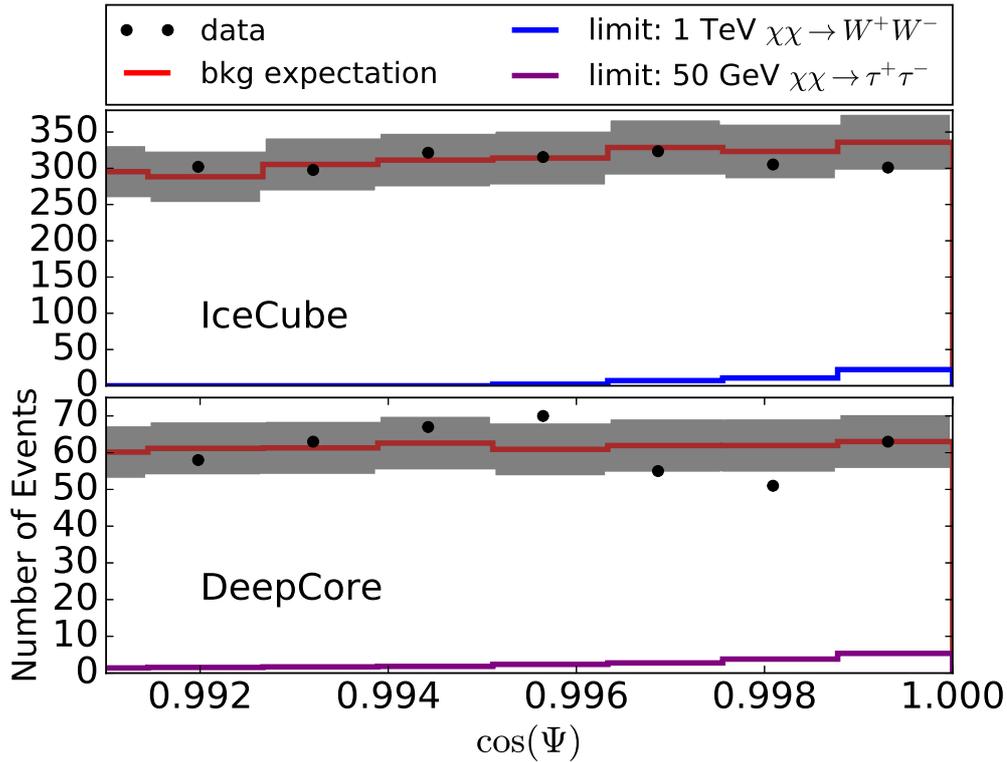}
  \caption{Distribution of cosine of the opening angles towards the Sun observed in events of the IceCube (top) and DeepCore (bottom) samples. The black dots represent the number of events reconstructed at the corresponding direction, the red lines are the average background expectations with the gray shaded regions corresponding to statistical uncertainties on the background expectations, while the blue lines indicate the events expected from WIMPs of masses {1\UNIT{TeV}} annihilating into \chW at {$2.84\mExp{19}$\UNIT{s$^{-1}$}} and {50\UNIT{GeV}} annihilating to {\chtau} at the rate of {$3.46\mExp{22}$\UNIT{s$^{-1}$}} respectively, the present upper limits. The analysis method employed is unbinned in direction, consequently the binning employed in this figure is for indicative purposes only.}
  \label{fig:EventDirectionSun}
\end{figure*}

Background levels are estimated in this analysis method using data randomized in right ascension (see Fig.~\ref{fig:EventDirectionSun}) and so are, by construction, free of significant systematic uncertainties. A previous study of the signal uncertainties on data from the 79-string configuration of IceCube~\cite{MatthiasPaper} concluded that the following sources of uncertainty are intrinsic to the signal simulation (percentage impact on sensitivity in parenthesis):
\begin{enumerate}
 \item Neutrino-nucleon cross sections (7\% at {$m_\chi<$35\UNIT{GeV}} down to 3.5\% for $m_\chi>$100\UNIT{GeV})
 \item Uncertainties in neutrino oscillation parameters (6\%)
 \item Uncertainties in muon propagation in ice ($<$1\%),
\end{enumerate}
while the following sources dominate the detection process
\begin{enumerate}
 \item Absolute DOM efficiency
 \item Photon propagation in ice (absorption and scattering).
\end{enumerate}
Since the first class of sources of uncertainties, direct inputs to the \texttt{WimpSim} signal generator that mostly affect the signal flux normalisations, have not significantly changed with respect to the study in Ref.~\cite{MatthiasPaper}, we assume them to be similar. The second class of sources of uncertainties also include effects altering the signal's apparent spectral composition in the detection process and are thus more important to this analysis which is sensitive to the reconstructed event energy.

To study the effect of the absolute DOM efficiency as well as the absorption and scattering properties of the ice, a set of signal simulations were generated for one year of data by individually varying each quantities by $\pm$10\% from the baseline value for certain benchmark signals of interest.

The percentage impact of these variations on the muon flux $\bar{\Phi}_{\mu+\bar{\mu}}$, which can be converted to the other quantities in Table~\ref{tab:WIMPresults}, are summarized in Table~\ref{tab:Systematicsresults}. The percentage impact of the uncertainties on neutrino-nucleon cross sections, neutrino oscillations and muon propagation in ice are taken from~\cite{MatthiasPaper} and summed in quadrature to the ones from uncertainties in DOM efficiency and ice optical properties to obtain the total systematic uncertainty.

\begin{table}[]
  \centering
  \caption[Systematics]{Systematics summary stating uncertainties in the detection process. The last column states the total uncertainty estimate, which is the largest sum in quadrature of the individual uncertainties.}
  \label{tab:Systematicsresults}
  \begin{tabular}{rc|cc||c}
    \toprule
    \specialcell{$m_{\chi}$\\(GeV)} 
    & \specialcell{annih.\\channel} 
    & \specialcell{absolute\\DOM eff (\%)} 
    & \specialcell{photon prop.\\in ice (\%)} 
    & \specialcell{total \\(\%)}\\
    \midrule
    20 & \chtau & $-$11/+29 & $-$13/+18 & 35 \\ 
    50 & \chtau & $-$8/+23 & $-$9/+13 & 29 \\ 
    100 & \chW & $-$9/+19 & $-$9/+11 & 23 \\ 
    500 & \chb & $-$7/+11 & $-$8/+7 & 15 \\ 
    1000 & \chW & $-$6/+9 & $-$6/+4 & 12 \\
    \bottomrule
  \end{tabular}
\end{table}

\section{Results}\label{sec:results}

No significant excess of events over the expected background was found in the direction of the Sun, allowing us to set limits on the neutrino flux from the Sun in the GeV--TeV range. Assuming a conservative local DM density of {0.3\UNIT{GeV/cm$^3$}}~\cite{LewinSmith}, a standard Maxwellian halo velocity distribution and the Standard Solar Model, this limit can also be interpreted as a limit on the WIMP-proton scattering cross section. Table~\ref{tab:WIMPresults} summarizes the best fit number of signal events and the upper limit on the muon and neutrino fluxes, as well as spin-dependent and spin-independent WIMP-proton scattering cross sections. 

\begin{figure*}[]\centering
  \includegraphics[width=0.9\linewidth]{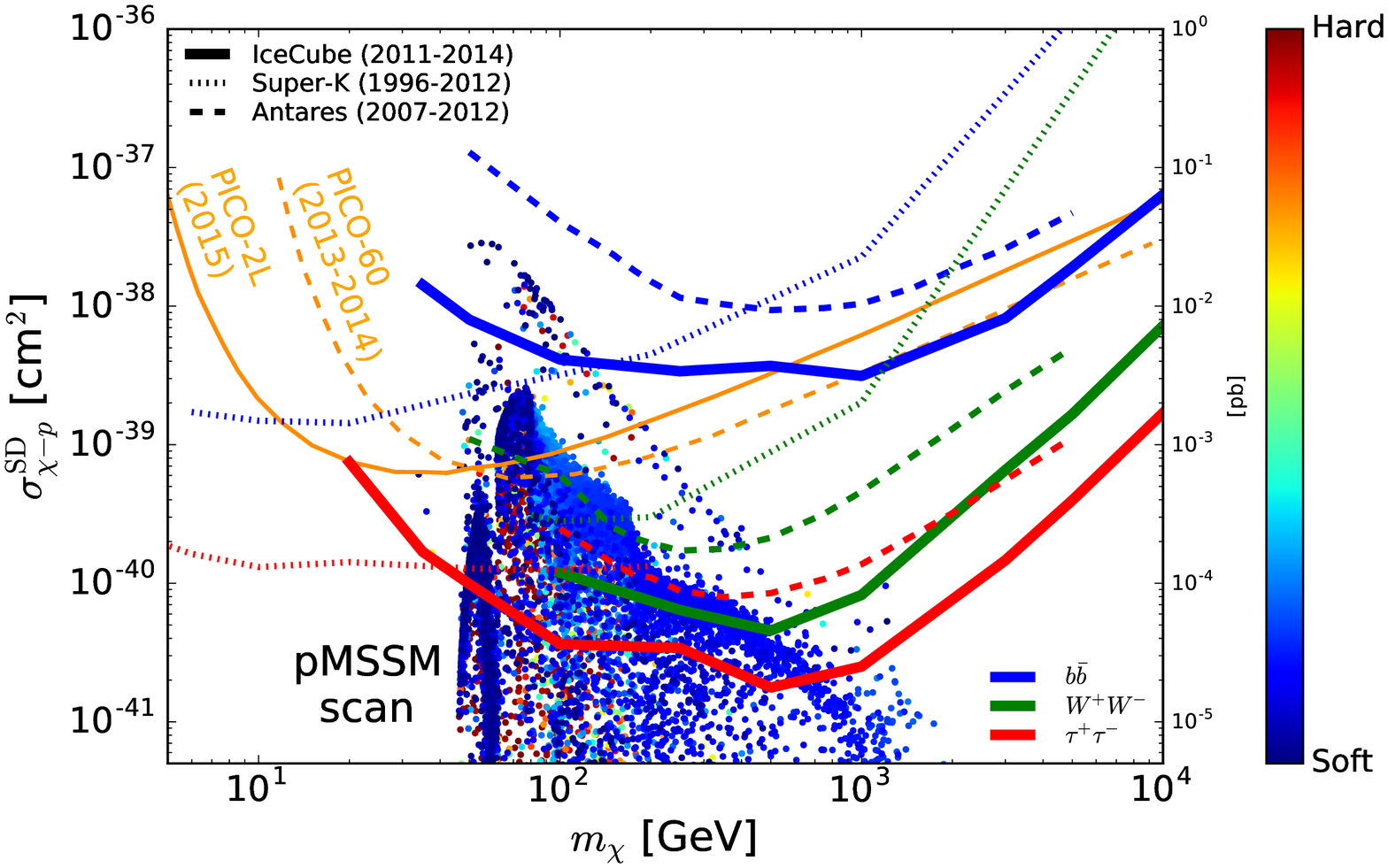}
  \caption{Limits on $\sigma^{\mathrm{SD}}_{\chi\mathrm{-}p}$, compared to results from other neutrino detectors and direct detection experiments~\cite{ANTARES, SuperK, PICO1, PICO2}. The IceCube limits have been scaled up to the upper edge of the total systematic uncertainty band. The colored points correspond to models from a scan of the pMSSM described in Section \ref{sec:concl} and are shown color coded by the `hardness' of the resultant neutrino spectrum. Points close to the red end of the spectrum annihilate predominantly into harder channels such as \chtau and can hence be excluded by the IceCube red line.
  }  
  \label{fig:SolarWIMPResults1}
\end{figure*}

\begin{figure*}[]\centering
  \includegraphics[width=0.9\linewidth]{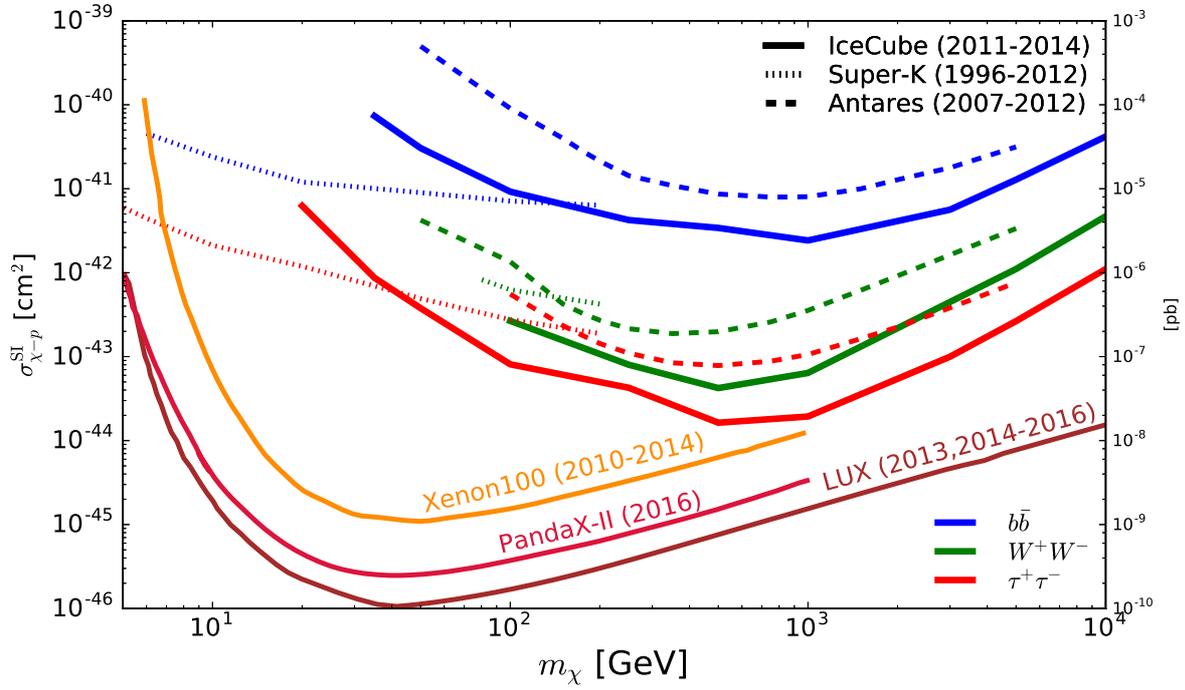}
  \caption{Limits on $\sigma^{\mathrm{SI}}_{\chi\mathrm{-}p}$, compared to results from other neutrino detectors and direct detection experiments~\cite{ANTARES, SuperK, LUX, PandaII, Xenon100}. The IceCube limits include the systematic uncertainties.}  
  \label{fig:SolarWIMPResults2}
\end{figure*}

\begin{table*}[]
  \centering
  \caption[Summary of Results]{{P-values} and 90\%~C.L.\ upper limits on the number of signal events within the two samples in $\sim$532\,days of livetime, corresponding to three years of operation of IceCube-DeepCore in its final configuration. The average effective volumes over the three years are also provided, as well as upper limits on the muon flux, annihilation rate, and the spin-dependent and spin-independent WIMP-proton scattering cross sections.}
  \label{tab:WIMPresults}
  \begin{tabular}{|r|c|c|c|c|c|c|c|c|c|c|}
    \toprule
    \specialcell{$m_{\chi}$\\(GeV)} &
    \specialcell{annih.\\channel} &
    \specialcell{dataset} &
    \specialcell{p-value\\{\%}} &
    $n_{\mathrm{s}}^{\mathrm{90\%C.L.}}$ &
    \specialcell{$V_{\mathrm{eff}}$\\(km$^3$)} &
    \specialcell{$\bar{\Phi}_{\mu+\bar{\mu}}$\\ (km$^{-2}$yr$^{-1}$)} &
    \specialcell{$\Phi_{\mu+\bar{\mu}}^{\mathrm{90\%C.L.}}$\\(km$^{-2}$yr$^{-1}$)} &
    \specialcell{$\Gamma_{\chi\chi \to \mathrm{SM}}^{\mathrm{90\%C.L.}}$\\(s$^{-1}$) } &
    \specialcell{$\sigma_{\mathrm{SD}}^{\mathrm{90\%C.L.}}$\\(pb)} &
    \specialcell{$\sigma_{\mathrm{SI}}^{90\%\mathrm{C.L.}}$\\(pb)}\\
    \midrule
    20    & \chtau  & DC    & $>$50 & 97.2 & 4.40e-04 & 1.58e+05 & 1.52e+05 & 9.19e+23 & 4.85e-04 & 4.06e-06 \\
    \hline                                                                                           
    35    & \chb      & DC    & $>$50 & 96.8 & 2.79e-04 & 2.44e+05 & 2.38e+05 & 7.39e+24 & 9.25e-03 & 4.77e-05 \\
    35    & \chtau  & DC    & $>$50 & 59.1 & 1.26e-03 & 3.33e+04 & 3.22e+04 & 1.08e+23 & 1.35e-04 & 6.95e-07 \\
    \hline                                                                                           
    50    & \chb      & DC    & $>$50 & 87.3 & 4.71e-04 & 1.29e+05 & 1.27e+05 & 2.79e+24 & 6.39e-03 & 2.44e-05 \\
    50    & \chtau  & DC    & 48.4  & 48.9 & 2.31e-03 & 1.39e+04 & 1.45e+04 & 3.46e+22 & 7.90e-05 & 3.02e-07 \\
    \hline                                                                                        
    100   & \chb      & DC    & 46.1  & 65.2 & 1.39e-03 & 3.05e+04 & 3.22e+04 & 4.09e+23 & 3.29e-03 & 7.38e-06 \\
    100   & \chW        & DC    & 34.7  & 36.1 & 6.64e-03 & 2.81e+03 & 3.73e+03 & 1.18e+22 & 9.52e-05 & 2.13e-07 \\
    100   & \chtau  & DC    & 31.3  & 37.6 & 9.40e-03 & 2.13e+03 & 2.75e+03 & 3.60e+21 & 2.91e-05 & 6.48e-08 \\
    \hline                                                                                            
    250  & \chb       & DC+IC & 28.2  & 55.1 & 4.42e-03 & 6.79e+03 & 8.57e+03 & 5.96e+22 & 2.80e-03 & 3.50e-06 \\
    250  & \chW         & DC+IC & 39.8  & 64.7 & 7.38e-02 & 5.03e+02 & 6.02e+02 & 1.13e+21 & 5.30e-05 & 6.62e-08 \\
    250  & \chtau   & DC+IC & 42.1  & 90.6 & 7.20e-02 & 7.74e+02 & 8.64e+02 & 5.99e+20 & 2.82e-05 & 3.52e-08 \\
    \hline                                  
    500   & \chb      & DC+IC & 46.1  & 75.6 & 1.54e-02 & 3.04e+03 & 3.37e+03 & 1.66e+22 & 3.06e-03 & 2.82e-06 \\
    500   & \chW        & IC    & 39.3  & 36.0 & 1.87e-01 & 4.04e+01 & 5.53e+01 & 2.04e+20 & 3.76e-05 & 3.49e-08 \\
    500   & \chtau  & IC    & 38.7  & 45.1 & 1.95e-01 & 4.71e+01 & 5.93e+01 & 7.96e+19 & 1.46e-05 & 1.35e-08 \\
    \hline                                                   
    1000  & \chb      & IC    & 37.2  & 43.1 & 3.24e-02 & 1.30e+02 & 1.55e+02 & 3.56e+21 & 2.59e-03 & 2.00e-06 \\
    1000  & \chW        & IC    & 48.9  & 24.6 & 2.67e-01 & 3.06e+01 & 3.31e+01 & 9.34e+19 & 6.80e-05 & 5.28e-08 \\
    1000  & \chtau  & IC    & 46.5  & 28.6 & 2.86e-01 & 3.30e+01 & 3.46e+01 & 2.84e+19 & 2.07e-05 & 1.60e-08 \\
    \hline                                                   
    3000  & \chb      & IC    & 48.2  & 32.1 & 6.62e-02 & 7.29e+01 & 7.56e+01 & 1.04e+21 & 6.76e-03 & 4.65e-06 \\
    3000  & \chW        & IC    & 49.6  & 23.1 & 2.86e-01 & 3.07e+01 & 3.13e+01 & 8.33e+19 & 5.42e-04 & 3.70e-07 \\
    3000  & \chtau  & IC    & 49.4  & 21.1 & 2.92e-01 & 2.85e+01 & 2.90e+01 & 1.85e+19 & 1.21e-04 & 8.25e-08 \\
    \hline                                                   
    5000  & \chb      & IC    & 49.1  & 33.7 & 7.72e-02 & 7.11e+01 & 7.24e+01 & 8.74e+20 & 1.58e-02 & 1.06e-05 \\
    5000  & \chW        & IC    & 49.8  & 22.4 & 3.09e-01 & 2.78e+01 & 2.84e+01 & 7.59e+19 & 1.37e-03 & 9.14e-07 \\
    5000  & \chtau  & IC    & 49.8  & 22.3 & 3.10e-01 & 2.86e+01 & 2.93e+01 & 1.82e+19 & 3.28e-04 & 2.19e-07 \\
    \hline                                                   
    10000 & \chb      & IC    & 49.8  & 32.5 & 8.26e-02 & 6.74e+01 & 6.87e+01 & 7.31e+20 & 5.27e-02 & 3.46e-05 \\
    10000 & \chW        & IC    & $>$50 & 25.2 & 3.18e-01 & 3.08e+01 & 3.11e+01 & 8.26e+19 & 5.96e-03 & 3.88e-06 \\
    10000 & \chtau  & IC    & $>$50 & 25.0 & 3.19e-01 & 3.18e+01 & 3.21e+01 & 1.94e+19 & 1.40e-03 & 9.11e-07 \\
    \bottomrule
  \end{tabular}
\end{table*}

\section{Conclusions and Interpretations}\label{sec:concl}
For spin-dependent WIMP-proton scattering, IceCube limits are the most competitive in the region above {$\sim$80\UNIT{GeV}} (Fig.~\ref{fig:SolarWIMPResults1}). The constraints on spin-independent scattering (Fig.~\ref{fig:SolarWIMPResults2}) from this search are complementary to the limits from direct detection. Even though these tend to be significantly stronger, they are subject to different uncertainties from the nuclear scattering process and from astrophysics. Limits have improved by a factor of $\sim$2 to 4 with respect to previous IceCube analyses~\cite{MatthiasPaper, IC79ImprovedPaper}. While these constraints explicitly assume equilibrium between capture and annihilation in the Sun, for the natural scale of $\left<\sigma_A v\right>$ {$\sim 3 \times 10^{-26}$\UNIT{cm$^3$s$^{-1}$}} \cite{WIMPnaturalscale}, the time required for equilibrium to be achieved becomes as large as the age of the Sun only for $\sigma^{\mathrm{SD}}_{\chi\mathrm{-}p}$ as low as {$10^{-43}$\UNIT{cm$^2$}}, making this a very reasonable assumption~\cite{RameezThesis}. The uncertainties on these limits due to uncertainties in velocity distributions of DM have been quantified in Ref.~\cite{CarstenChoiPaper} and do not exceed $\sim$50\%. The study also concludes that these limits are conservative with respect to the possible existence of a dark-disk~\cite{DarkDisk}, since a population of DM particles with lower velocities in the disk will enhance the capture rate. For a dark disk contributing an additional 25\% to the local DM density, co-rotating with the visible stellar disk with no lag in velocity and a velocity dispersion of {$\sigma=50$\UNIT{km/s}}, the spin-dependent capture rate is boosted by a factor of $\sim$20 at high DM masses, improving the constraint on the spin-dependent cross section by a corresponding amount. 

As demonstrated in Fig.~\ref{fig:SolarWIMPResults1}, these constraints exclude some models corresponding to neutralinos from a scan of $\sim$500 million points in the 19 parameter realization of the phenomenological minimally super-symmetric standard model (pMSSM)~\cite{Djouadi:1998di,Berger:2008cq} performed using \texttt{micrOMEGAs}~\cite{micromegas} with logarithmically distributed priors (chosen to preferentially populate low mass, high $\sigma^{\mathrm{SD}}_{\chi\mathrm{-}p}$ models) on the mass parameters typically in the range {50--10000\UNIT{GeV/c$^2$}}. The points were required to yield a relic dark matter density consistent with PLANCK measurements~\cite{Ade:2015xua} and a Higgs mass within the currently known uncertainty range~\cite{Aad:2015zhl}, in addition to being consistent with recent measurements of the $B_{0}^{s} \to \mu^{+} \mu^{-}$ branching ratio~\cite{CMS:2014xfa} and the CKM matrix element $V_{ub}$~\cite{Aaij:2015bfa}. Further details are given in~\cite{klausphd}. 

Beyond the WIMP paradigm, this search is sensitive to any scenario with a DM particle in the {20\UNIT{GeV}} to {10\UNIT{TeV}} mass range that can scatter off nuclei sufficiently strongly to cause an over-density at the center of the Sun, and can annihilate to produce neutrinos as primary or secondary products. Some specific scenarios have been considered in~\cite{Zdashpaper}. Scenarios where the DM-nucleon scattering is velocity or momentum dependent and hence suppressed at non relativistic energies are also of particular interest~\cite{CatanoEFT}. In these scenarios, this search can be significantly more powerful than direct detection constraints, due to the fact that capture in the Sun for a DM-nucleus interaction that depends on the spin of the nucleus is dominated by scattering off light nuclei, while direct detection experiments on Earth rely on significantly heavier nuclear targets. Theories with DM candidates that interact very differently with protons and neutrons are also better constrained by this search, which relies on DM scattering off a democratic distribution of various nuclei present in the Sun, each with a different neutron-proton ratio. This is in contrast to direct detection experiments which often rely on a single target nucleus specimen~\cite{IsoViol, IsoViol2}.

\input{bibliography}
\input{acknowledgements}

\end{document}

%% file: ICauthors.tex
\author{IceCube Collaboration: M.~G.~Aartsen\thanksref{Adelaide}
\and M.~Ackermann\thanksref{Zeuthen}
\and J.~Adams\thanksref{Christchurch}
\and J.~A.~Aguilar\thanksref{BrusselsLibre}
\and M.~Ahlers\thanksref{MadisonPAC}
\and M.~Ahrens\thanksref{StockholmOKC}
\and D.~Altmann\thanksref{Erlangen}
\and K.~Andeen\thanksref{Marquette}
\and T.~Anderson\thanksref{PennPhys}
\and I.~Ansseau\thanksref{BrusselsLibre}
% ...
}
\authorrunning{IceCube Collaboration}
\thankstext{a}{Earthquake Research Institute, University of Tokyo, Bunkyo, Tokyo 113-0032, Japan}
\institute{
% ...
Department of Physics, University of Adelaide, Adelaide, 5005, Australia \label{Adelaide}
% ...
\and Universit\'e Libre de Bruxelles, Science Faculty CP230, B-1050 Brussels, Belgium \label{BrusselsLibre}
% ...
\and Dept.~of Physics and Astronomy, University of Canterbury, Private Bag 4800, Christchurch, New Zealand \label{Christchurch}
% ...
\and Erlangen Centre for Astroparticle Physics, Friedrich-Alexander-Universit\"at Erlangen-N\"urnberg, D-91058 Erlangen, Germany \label{Erlangen}
% ...
\and Dept.~of Physics and Wisconsin IceCube Particle Astrophysics Center, University of Wisconsin, Madison, WI 53706, USA \label{MadisonPAC}
% ...
\and Department of Physics, Marquette University, Milwaukee, WI, 53201, USA \label{Marquette}
% ...
\and Oskar Klein Centre and Dept.~of Physics, Stockholm University, SE-10691 Stockholm, Sweden \label{StockholmOKC}
% ...
\and Dept.~of Physics, Pennsylvania State University, University Park, PA 16802, USA \label{PennPhys}
% ...
\and DESY, D-15735 Zeuthen, Germany \label{Zeuthen}
} 
\thankstext{corr_a}{e-mail: mohamed.rameez@nbi.ku.dk}\thankstext{corr_b}{e-mail: marcel.zoll.physics@gmail.com} 

%% file: acknowledgements.tex
\begin{acknowledgements}
We acknowledge the support from the following agencies:
U.S.\ National Science Foundation-Office of Polar Programs,
U.S.\ National Science Foundation-Physics Division,
University of Wisconsin Alumni Research Foundation,
the Grid Laboratory Of Wisconsin (GLOW) grid infrastructure at the University of Wisconsin - Madison, the Open Science Grid (OSG) grid infrastructure;
U.S.\ Department of Energy, and National Energy Research Scientific Computing Center,
the Louisiana Optical Network Initiative (LONI) grid computing resources;
Natural Sciences and Engineering Research Council of Canada,
WestGrid and Compute/Calcul Canada;
Swedish Research Council,
Swedish Polar Research Secretariat,
Swedish National Infrastructure for Computing (SNIC),
and Knut and Alice Wallenberg Foundation, Sweden;
German Ministry for Education and Research (BMBF),
Deutsche Forschungsgemeinschaft (DFG),
Helmholtz Alliance for Astroparticle Physics (HAP),
Research Department of Plasmas with Complex Interactions (Bochum), Germany;
Fund for Scientific Research (FNRS-FWO),
FWO Odysseus programme,
Flanders Institute to encourage scientific and technological research in industry (IWT),
Belgian Federal Science Policy Office (Belspo);
University of Oxford, United Kingdom;
Marsden Fund, New Zealand;
Australian Research Council;
Japan Society for Promotion of Science (JSPS);
the Swiss National Science Foundation (SNSF), Switzerland;
National Research Foundation of Korea (NRF);
Villum Fonden, Danish National Research Foundation (DNRF), Denmark

\end{acknowledgements}